\documentstyle[twoside,fleqn,espcrc2,epsf]{article}

\newcommand{\AmS}{{\protect\the\textfont2
  A\kern-.1667em\lower.5ex\hbox{M}\kern-.125emS}}

\title{Probability of Incipient Spanning Clusters 
in Critical Two-Dimensional Percolation} 

\author{L.N. Shchur
        \thanks{E-mail address: lev@itp.ac.ru.
        Partialy supported by grants RFBR 96-02-18168 and 
        96-07-89266, NWO 07-210 and INTAS 93-211}
        and 
        S.S. Kosyakov \\ \vspace{9pt}
        {Landau Institute for Theoretical Physics,
        142432 Chernogolovka, Russia}}
       
\begin{document}

\begin{abstract}
The probability of simultaneous occurence of at least $k$ spanning clusters
has been studied by Monte Carlo simulations on the 2D square lattice at
the bond percolation threshold $p_c=1/2$.
The calculated probabilities for free boundary conditions and 
periodic boundary conditions are in a very good coincidence with 
the exact formulae developed recently by Cardy.
\end{abstract}

\maketitle

\section{INTRODUCTION}

It was a common belief until a very recent time 
that on two-dimensional (2D) lattices at percolation threshold $p_c$
there exists exactly one percolation cluster \cite{Intro,Grimt}. 
New insight developed recently by Aizenman,
who proved \cite{Aiz1} that the number of Incipient Spanning Clusters 
(ISC) in 2D critical percolation can be larger than one,
and that the probability of at least $k$ separate clusters is bounded
$ A\; e^{-\alpha \  k^2} \le P_L(k) \le  e^{-\alpha' \  k^2} $,
where $\alpha$ and $\alpha'$ are constants and $L$ is a linear 
lattice size. 
We investigate by Monte-Carlo the number of spanning clusters 
in the critical bond percolation model on 2D square
lattices. Using simple finite-size scaling of probabilities on a self-dual
lattices of moderate size,
we have determined with a good accuracy
values of the probabilities 
$P(k) =  \lim_{L\rightarrow\infty} \  P_L(k)  \ $ for $k=1,2$ and $3$.

\section{LATTICE}

\begin{figure}[htb]
\epsfxsize=60mm
\epsffile{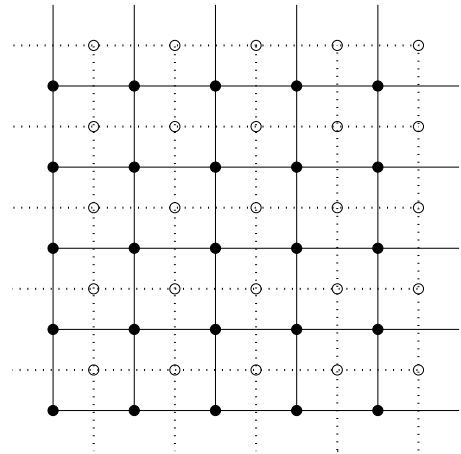}
\caption{Example of lattice with $L=5$.}
\label{fig:Lat}
\end{figure}

\begin{figure}[htb]
\vspace{7pt}
\setlength{\unitlength}{0.240900pt}
\ifx\plotpoint\undefined\newsavebox{\plotpoint}\fi
\sbox{\plotpoint}{\rule[-0.200pt]{0.400pt}{0.400pt}}%
\begin{picture}(900,720)(0,0)
\font\gnuplot=cmr10 at 10pt
\gnuplot
\sbox{\plotpoint}{\rule[-0.200pt]{0.400pt}{0.400pt}}%
\put(220.0,113.0){\rule[-0.200pt]{0.400pt}{140.686pt}}
\put(220.0,113.0){\rule[-0.200pt]{4.818pt}{0.400pt}}
\put(198,113){\makebox(0,0)[r]{6.4}}
\put(816.0,113.0){\rule[-0.200pt]{4.818pt}{0.400pt}}
\put(220.0,196.0){\rule[-0.200pt]{4.818pt}{0.400pt}}
\put(198,196){\makebox(0,0)[r]{6.6}}
\put(816.0,196.0){\rule[-0.200pt]{4.818pt}{0.400pt}}
\put(220.0,280.0){\rule[-0.200pt]{4.818pt}{0.400pt}}
\put(198,280){\makebox(0,0)[r]{6.8}}
\put(816.0,280.0){\rule[-0.200pt]{4.818pt}{0.400pt}}
\put(220.0,363.0){\rule[-0.200pt]{4.818pt}{0.400pt}}
\put(198,363){\makebox(0,0)[r]{7}}
\put(816.0,363.0){\rule[-0.200pt]{4.818pt}{0.400pt}}
\put(220.0,447.0){\rule[-0.200pt]{4.818pt}{0.400pt}}
\put(198,447){\makebox(0,0)[r]{7.2}}
\put(816.0,447.0){\rule[-0.200pt]{4.818pt}{0.400pt}}
\put(220.0,530.0){\rule[-0.200pt]{4.818pt}{0.400pt}}
\put(198,530){\makebox(0,0)[r]{7.4}}
\put(816.0,530.0){\rule[-0.200pt]{4.818pt}{0.400pt}}
\put(220.0,614.0){\rule[-0.200pt]{4.818pt}{0.400pt}}
\put(198,614){\makebox(0,0)[r]{7.6}}
\put(816.0,614.0){\rule[-0.200pt]{4.818pt}{0.400pt}}
\put(220.0,697.0){\rule[-0.200pt]{4.818pt}{0.400pt}}
\put(198,697){\makebox(0,0)[r]{7.8}}
\put(816.0,697.0){\rule[-0.200pt]{4.818pt}{0.400pt}}
\put(220.0,113.0){\rule[-0.200pt]{0.400pt}{4.818pt}}
\put(220,68){\makebox(0,0){0}}
\put(220.0,677.0){\rule[-0.200pt]{0.400pt}{4.818pt}}
\put(401.0,113.0){\rule[-0.200pt]{0.400pt}{4.818pt}}
\put(401,68){\makebox(0,0){0.005}}
\put(401.0,677.0){\rule[-0.200pt]{0.400pt}{4.818pt}}
\put(582.0,113.0){\rule[-0.200pt]{0.400pt}{4.818pt}}
\put(582,68){\makebox(0,0){0.01}}
\put(582.0,677.0){\rule[-0.200pt]{0.400pt}{4.818pt}}
\put(764.0,113.0){\rule[-0.200pt]{0.400pt}{4.818pt}}
\put(764,68){\makebox(0,0){0.015}}
\put(764.0,677.0){\rule[-0.200pt]{0.400pt}{4.818pt}}
\put(220.0,113.0){\rule[-0.200pt]{148.394pt}{0.400pt}}
\put(836.0,113.0){\rule[-0.200pt]{0.400pt}{140.686pt}}
\put(220.0,697.0){\rule[-0.200pt]{148.394pt}{0.400pt}}
\put(95,750){\makebox(0,0){$P(k>1)\cdot 10^3$}}
\put(528,3){\makebox(0,0){$1/L^2$}}
\put(220.0,113.0){\rule[-0.200pt]{0.400pt}{140.686pt}}
\put(786,638){\raisebox{-.8pt}{\makebox(0,0){$\Diamond$}}}
\put(471,396){\raisebox{-.8pt}{\makebox(0,0){$\Diamond$}}}
\put(362,302){\raisebox{-.8pt}{\makebox(0,0){$\Diamond$}}}
\put(311,256){\raisebox{-.8pt}{\makebox(0,0){$\Diamond$}}}
\put(260,218){\raisebox{-.8pt}{\makebox(0,0){$\Diamond$}}}
\put(256,215){\raisebox{-.8pt}{\makebox(0,0){$\Diamond$}}}
\put(229,196){\raisebox{-.8pt}{\makebox(0,0){$\Diamond$}}}
\put(220,188){\usebox{\plotpoint}}
\multiput(220,188)(16.186,12.992){37}{\usebox{\plotpoint}}
\put(818,668){\usebox{\plotpoint}}
\end{picture}
\caption{Probability of more than one Incipient Spanning Cluster.}
\label{fig:P1}
\end{figure}
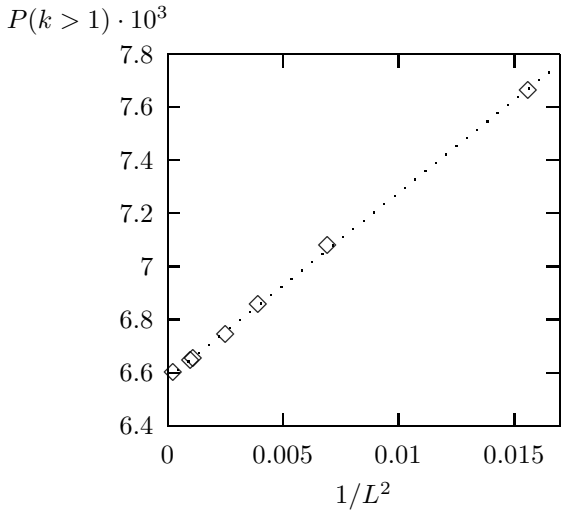

We use in simulations a rectangular square lattice with linear size $L$,
and exactly $L$ sites and $L$ bonds both horizontally and vertically.
For clarity the example of lattice is shown in Fig.~\ref{fig:Lat} together 
with the dual lattice. The dual lattice has the same number of sites and 
bonds as the original lattice. This gives the possibility to keep in the 
finite lattices some properties of infinite lattice. 
First of all, the  number of bonds is exactly twice
the number of sites. 
Second, the self-duality is valid for any finite lattice size. 
Third, the  horizontal and vertical directions are equivalent. 
Finally, one could expect trivial finite-size scaling for the
bond percolation on such lattices. Indeed, probabilities are proportional 
to the inverse system "volume", i.e. $\propto 1/L^2$ (Fig.~\ref{fig:P1} and 
Fig.~\ref{fig:P2}).

\section{ALGORITHM}

Each bond could be occupied with probability $q=1-p$ and closed with 
the probability $p$. Given the realization to each of $2\cdot L^2$ bonds to
be occupied formed the sample. Each sample could be
exactly decomposed in clusters of connected occupied bonds. 
We are interested in the spanning properties of such a clusters. Namely, 
what is the probability that a cluster connects the opposite borders of
square and what is the number of disjoint spanning clusters?
It should be noted that the variance of probabilities is independent from
the lattice size because probabilities are calculated as the expectation
values of corresponding indicator functions. 
So, we should keep the number of samples independent of the lattice
size and, therefore,  the only parameter which controls the
statistical errors is the number of samples $M$. Probabilities were
calculated averaging $M=10^8$ samples and the error bars were defined from 
$100$ bins, each bin being the average over $10^6$ samples.

\section{NUMERICAL RESULTS}

We calculate the crossing probabilities $\pi_{h_k}$ and $\pi_{v_k}$
for square lattices with linear sizes $L=8,12,16,20,30,32,64$ looking 
for all events up to $k=8$. See for details the more full text \cite{SK}.
On the Fig.~\ref{fig:P1} we plot the probability for simultaneous 
occurence of 
more than one spanning cluster $P_L(k>1) = \sum_{k>1} \pi_{h_k}$
versus the inverse system volume $1/L^2$.  A linear fit gives us the
limiting value of $P(k>1)=6.58\cdot 10^{-3}$ with error $\approx 3\cdot 
10^{-5}$.
\begin{figure}[htb]
\vspace{7pt}
\setlength{\unitlength}{0.240900pt}
\ifx\plotpoint\undefined\newsavebox{\plotpoint}\fi
\begin{picture}(900,720)(0,0)
\font\gnuplot=cmr10 at 10pt
\gnuplot
\sbox{\plotpoint}{\rule[-0.200pt]{0.400pt}{0.400pt}}%
\put(220.0,113.0){\rule[-0.200pt]{0.400pt}{140.686pt}}
\put(220.0,113.0){\rule[-0.200pt]{4.818pt}{0.400pt}}
\put(198,113){\makebox(0,0)[r]{1}}
\put(816.0,113.0){\rule[-0.200pt]{4.818pt}{0.400pt}}
\put(220.0,210.0){\rule[-0.200pt]{4.818pt}{0.400pt}}
\put(198,210){\makebox(0,0)[r]{1.5}}
\put(816.0,210.0){\rule[-0.200pt]{4.818pt}{0.400pt}}
\put(220.0,308.0){\rule[-0.200pt]{4.818pt}{0.400pt}}
\put(198,308){\makebox(0,0)[r]{2}}
\put(816.0,308.0){\rule[-0.200pt]{4.818pt}{0.400pt}}
\put(220.0,405.0){\rule[-0.200pt]{4.818pt}{0.400pt}}
\put(198,405){\makebox(0,0)[r]{2.5}}
\put(816.0,405.0){\rule[-0.200pt]{4.818pt}{0.400pt}}
\put(220.0,502.0){\rule[-0.200pt]{4.818pt}{0.400pt}}
\put(198,502){\makebox(0,0)[r]{3}}
\put(816.0,502.0){\rule[-0.200pt]{4.818pt}{0.400pt}}
\put(220.0,600.0){\rule[-0.200pt]{4.818pt}{0.400pt}}
\put(198,600){\makebox(0,0)[r]{3.5}}
\put(816.0,600.0){\rule[-0.200pt]{4.818pt}{0.400pt}}
\put(220.0,697.0){\rule[-0.200pt]{4.818pt}{0.400pt}}
\put(198,697){\makebox(0,0)[r]{4}}
\put(816.0,697.0){\rule[-0.200pt]{4.818pt}{0.400pt}}
\put(220.0,113.0){\rule[-0.200pt]{0.400pt}{4.818pt}}
\put(220,68){\makebox(0,0){0}}
\put(220.0,677.0){\rule[-0.200pt]{0.400pt}{4.818pt}}
\put(401.0,113.0){\rule[-0.200pt]{0.400pt}{4.818pt}}
\put(401,68){\makebox(0,0){0.005}}
\put(401.0,677.0){\rule[-0.200pt]{0.400pt}{4.818pt}}
\put(582.0,113.0){\rule[-0.200pt]{0.400pt}{4.818pt}}
\put(582,68){\makebox(0,0){0.01}}
\put(582.0,677.0){\rule[-0.200pt]{0.400pt}{4.818pt}}
\put(764.0,113.0){\rule[-0.200pt]{0.400pt}{4.818pt}}
\put(764,68){\makebox(0,0){0.015}}
\put(764.0,677.0){\rule[-0.200pt]{0.400pt}{4.818pt}}
\put(220.0,113.0){\rule[-0.200pt]{148.394pt}{0.400pt}}
\put(836.0,113.0){\rule[-0.200pt]{0.400pt}{140.686pt}}
\put(220.0,697.0){\rule[-0.200pt]{148.394pt}{0.400pt}}
\put(95,750){\makebox(0,0){$P(k>2)\cdot 10^6$}}
\put(528,3){\makebox(0,0){$1/L^2$}}
\put(220.0,113.0){\rule[-0.200pt]{0.400pt}{140.686pt}}
\put(786,637){\raisebox{-.8pt}{\makebox(0,0){$\Diamond$}}}
\put(471,373){\raisebox{-.8pt}{\makebox(0,0){$\Diamond$}}}
\put(362,284){\raisebox{-.8pt}{\makebox(0,0){$\Diamond$}}}
\put(311,276){\raisebox{-.8pt}{\makebox(0,0){$\Diamond$}}}
\put(260,214){\raisebox{-.8pt}{\makebox(0,0){$\Diamond$}}}
\put(256,239){\raisebox{-.8pt}{\makebox(0,0){$\Diamond$}}}
\put(229,195){\raisebox{-.8pt}{\makebox(0,0){$\Diamond$}}}
\put(220,191){\usebox{\plotpoint}}
\multiput(220,191)(16.584,12.480){37}{\usebox{\plotpoint}}
\put(818,641){\usebox{\plotpoint}}
\end{picture}
\caption{Probability of more than two Incipient Spanning Cluster.}
\label{fig:P2}
\end{figure}
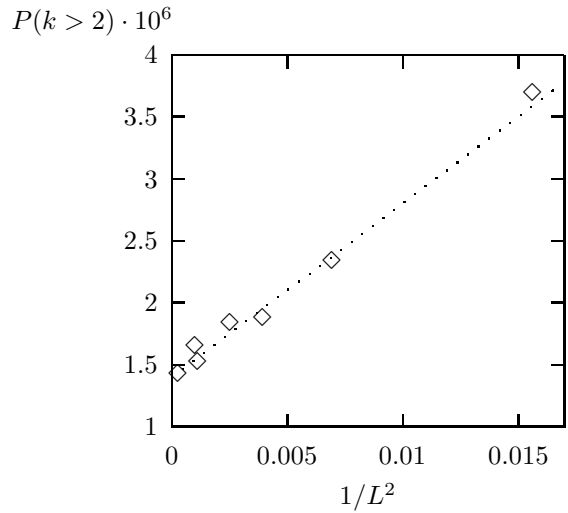
 Fig.~\ref{fig:P2} shows the dependence of the probability of more than two 
clusters versus $1/L^2$. The best linear fit gives the limiting value of
$P(k>2)=1.48\cdot 10^{-6}$ with uncertainty $2.1\cdot 10^{-7}$.

 Actually, we observed in computations mostly the events of up to
three simultaneous spanning clusters.  We simulated at total 
about $10^{10}$ samples of different sizes $8 \le L \le 64$
and only one sample with $4$ spanning clusters was detected. This event
clearly not contradicts with our estimate \cite{SK} for $P(4)\approx 
10^{-11}$. Single event is probable as one part in ten.

 A best linear fit to the logarithm of $P(k)$ versus $k^2$ gives $\alpha 
\propto 1.61(7)$ and $A\propto 3\pm1.5$. 

The crossing probability with periodic boundary conditions (PBC) is known
to be larger and the value $0.63665(8)$ is computed in \cite{AH-PRE}.
We estimate in computer simulations the probability of
disjoint ISC clusters to be $P(k>1) \approx 2.0(4) \ 10^{-3}$ and
$P(k>2) \approx 1.4(5)\ 10^{-7}$ which are even smaller than for the
case of free boundary conditions (FBC) considered by us.
The finite size scaling of PBC is more complicated \cite{AH-PRE}
and fit of the results for only four lattice sizes (8,16,32,64) we
simulated gives less accurate limiting values in comparison with the
one for FBC. Surprisingly for us, the linear fit of logarithm of $P(k)$
versus $k^2$ gives better accuracy 
then for the case of FBC: $\alpha_{PBC} \approx 1.915(1)$ and 
$A\approx 4.26(3)$. The different behaviour of probabilities will be 
explained in the next section.

\section{COMPARISON WITH EXACT RESULT}

Quite recently, John Cardy,
using methods of conformal field theory, conjectured an exact form 
\cite{Cardy} of probability of $k$ distinct Incipient Spanning Clusters in
the limit of large aspect ratio $r$ for rectangle (in our notations, FBC -
free boundary conditions) and cylinder (PBC - periodic boundary 
conditions). It is possible to compare the data of our paper with the
ratio, developed from formulae (2) and (3) of Cardy preprint
$$
\frac{log P_{FBC}(k)}{log P_{PBC}(k)}=\frac{k(k-\frac12) }
{k^2-\frac14 },
$$
which is $\frac45 =0.8$ for $k=2$ and $\frac67 \approx 0.857$ for $k=3$.
Our data gives $0.808(10)$ and $0.851(20)$, correspondingly. 
The discussion of that close agreement can be found in Cardy's preprint.
The difference in the $k$-dependence explains why our PBC data fits more
closely to $k^2$ than the FBC ones.
\begin{figure}[htb]
\vspace{7pt}
\setlength{\unitlength}{0.240900pt}
\ifx\plotpoint\undefined\newsavebox{\plotpoint}\fi
\sbox{\plotpoint}{\rule[-0.200pt]{0.400pt}{0.400pt}}%
\begin{picture}(900,720)(0,0)
\font\gnuplot=cmr10 at 10pt
\gnuplot
\sbox{\plotpoint}{\rule[-0.200pt]{0.400pt}{0.400pt}}%
\put(220.0,113.0){\rule[-0.200pt]{0.400pt}{140.686pt}}
\put(220.0,113.0){\rule[-0.200pt]{4.818pt}{0.400pt}}
\put(198,113){\makebox(0,0)[r]{1e-08}}
\put(816.0,113.0){\rule[-0.200pt]{4.818pt}{0.400pt}}
\put(220.0,135.0){\rule[-0.200pt]{2.409pt}{0.400pt}}
\put(826.0,135.0){\rule[-0.200pt]{2.409pt}{0.400pt}}
\put(220.0,164.0){\rule[-0.200pt]{2.409pt}{0.400pt}}
\put(826.0,164.0){\rule[-0.200pt]{2.409pt}{0.400pt}}
\put(220.0,179.0){\rule[-0.200pt]{2.409pt}{0.400pt}}
\put(826.0,179.0){\rule[-0.200pt]{2.409pt}{0.400pt}}
\put(220.0,186.0){\rule[-0.200pt]{4.818pt}{0.400pt}}
\put(198,186){\makebox(0,0)[r]{1e-07}}
\put(816.0,186.0){\rule[-0.200pt]{4.818pt}{0.400pt}}
\put(220.0,208.0){\rule[-0.200pt]{2.409pt}{0.400pt}}
\put(826.0,208.0){\rule[-0.200pt]{2.409pt}{0.400pt}}
\put(220.0,237.0){\rule[-0.200pt]{2.409pt}{0.400pt}}
\put(826.0,237.0){\rule[-0.200pt]{2.409pt}{0.400pt}}
\put(220.0,252.0){\rule[-0.200pt]{2.409pt}{0.400pt}}
\put(826.0,252.0){\rule[-0.200pt]{2.409pt}{0.400pt}}
\put(220.0,259.0){\rule[-0.200pt]{4.818pt}{0.400pt}}
\put(198,259){\makebox(0,0)[r]{1e-06}}
\put(816.0,259.0){\rule[-0.200pt]{4.818pt}{0.400pt}}
\put(220.0,281.0){\rule[-0.200pt]{2.409pt}{0.400pt}}
\put(826.0,281.0){\rule[-0.200pt]{2.409pt}{0.400pt}}
\put(220.0,310.0){\rule[-0.200pt]{2.409pt}{0.400pt}}
\put(826.0,310.0){\rule[-0.200pt]{2.409pt}{0.400pt}}
\put(220.0,325.0){\rule[-0.200pt]{2.409pt}{0.400pt}}
\put(826.0,325.0){\rule[-0.200pt]{2.409pt}{0.400pt}}
\put(220.0,332.0){\rule[-0.200pt]{4.818pt}{0.400pt}}
\put(198,332){\makebox(0,0)[r]{1e-05}}
\put(816.0,332.0){\rule[-0.200pt]{4.818pt}{0.400pt}}
\put(220.0,354.0){\rule[-0.200pt]{2.409pt}{0.400pt}}
\put(826.0,354.0){\rule[-0.200pt]{2.409pt}{0.400pt}}
\put(220.0,383.0){\rule[-0.200pt]{2.409pt}{0.400pt}}
\put(826.0,383.0){\rule[-0.200pt]{2.409pt}{0.400pt}}
\put(220.0,398.0){\rule[-0.200pt]{2.409pt}{0.400pt}}
\put(826.0,398.0){\rule[-0.200pt]{2.409pt}{0.400pt}}
\put(220.0,405.0){\rule[-0.200pt]{4.818pt}{0.400pt}}
\put(198,405){\makebox(0,0)[r]{0.0001}}
\put(816.0,405.0){\rule[-0.200pt]{4.818pt}{0.400pt}}
\put(220.0,427.0){\rule[-0.200pt]{2.409pt}{0.400pt}}
\put(826.0,427.0){\rule[-0.200pt]{2.409pt}{0.400pt}}
\put(220.0,456.0){\rule[-0.200pt]{2.409pt}{0.400pt}}
\put(826.0,456.0){\rule[-0.200pt]{2.409pt}{0.400pt}}
\put(220.0,471.0){\rule[-0.200pt]{2.409pt}{0.400pt}}
\put(826.0,471.0){\rule[-0.200pt]{2.409pt}{0.400pt}}
\put(220.0,478.0){\rule[-0.200pt]{4.818pt}{0.400pt}}
\put(198,478){\makebox(0,0)[r]{0.001}}
\put(816.0,478.0){\rule[-0.200pt]{4.818pt}{0.400pt}}
\put(220.0,500.0){\rule[-0.200pt]{2.409pt}{0.400pt}}
\put(826.0,500.0){\rule[-0.200pt]{2.409pt}{0.400pt}}
\put(220.0,529.0){\rule[-0.200pt]{2.409pt}{0.400pt}}
\put(826.0,529.0){\rule[-0.200pt]{2.409pt}{0.400pt}}
\put(220.0,544.0){\rule[-0.200pt]{2.409pt}{0.400pt}}
\put(826.0,544.0){\rule[-0.200pt]{2.409pt}{0.400pt}}
\put(220.0,551.0){\rule[-0.200pt]{4.818pt}{0.400pt}}
\put(198,551){\makebox(0,0)[r]{0.01}}
\put(816.0,551.0){\rule[-0.200pt]{4.818pt}{0.400pt}}
\put(220.0,573.0){\rule[-0.200pt]{2.409pt}{0.400pt}}
\put(826.0,573.0){\rule[-0.200pt]{2.409pt}{0.400pt}}
\put(220.0,602.0){\rule[-0.200pt]{2.409pt}{0.400pt}}
\put(826.0,602.0){\rule[-0.200pt]{2.409pt}{0.400pt}}
\put(220.0,617.0){\rule[-0.200pt]{2.409pt}{0.400pt}}
\put(826.0,617.0){\rule[-0.200pt]{2.409pt}{0.400pt}}
\put(220.0,624.0){\rule[-0.200pt]{4.818pt}{0.400pt}}
\put(198,624){\makebox(0,0)[r]{0.1}}
\put(816.0,624.0){\rule[-0.200pt]{4.818pt}{0.400pt}}
\put(220.0,646.0){\rule[-0.200pt]{2.409pt}{0.400pt}}
\put(826.0,646.0){\rule[-0.200pt]{2.409pt}{0.400pt}}
\put(220.0,675.0){\rule[-0.200pt]{2.409pt}{0.400pt}}
\put(826.0,675.0){\rule[-0.200pt]{2.409pt}{0.400pt}}
\put(220.0,690.0){\rule[-0.200pt]{2.409pt}{0.400pt}}
\put(826.0,690.0){\rule[-0.200pt]{2.409pt}{0.400pt}}
\put(220.0,697.0){\rule[-0.200pt]{4.818pt}{0.400pt}}
\put(198,697){\makebox(0,0)[r]{1}}
\put(816.0,697.0){\rule[-0.200pt]{4.818pt}{0.400pt}}
\put(220.0,113.0){\rule[-0.200pt]{0.400pt}{4.818pt}}
\put(220,68){\makebox(0,0){0}}
\put(220.0,677.0){\rule[-0.200pt]{0.400pt}{4.818pt}}
\put(343.0,113.0){\rule[-0.200pt]{0.400pt}{4.818pt}}
\put(343,68){\makebox(0,0){2}}
\put(343.0,677.0){\rule[-0.200pt]{0.400pt}{4.818pt}}
\put(466.0,113.0){\rule[-0.200pt]{0.400pt}{4.818pt}}
\put(466,68){\makebox(0,0){4}}
\put(466.0,677.0){\rule[-0.200pt]{0.400pt}{4.818pt}}
\put(590.0,113.0){\rule[-0.200pt]{0.400pt}{4.818pt}}
\put(590,68){\makebox(0,0){6}}
\put(590.0,677.0){\rule[-0.200pt]{0.400pt}{4.818pt}}
\put(713.0,113.0){\rule[-0.200pt]{0.400pt}{4.818pt}}
\put(713,68){\makebox(0,0){8}}
\put(713.0,677.0){\rule[-0.200pt]{0.400pt}{4.818pt}}
\put(836.0,113.0){\rule[-0.200pt]{0.400pt}{4.818pt}}
\put(836,68){\makebox(0,0){10}}
\put(836.0,677.0){\rule[-0.200pt]{0.400pt}{4.818pt}}
\put(220.0,113.0){\rule[-0.200pt]{148.394pt}{0.400pt}}
\put(836.0,113.0){\rule[-0.200pt]{0.400pt}{140.686pt}}
\put(220.0,697.0){\rule[-0.200pt]{148.394pt}{0.400pt}}
\put(95,750){\makebox(0,0){log $P(k)$}}
\put(528,3){\makebox(0,0){$X$}}
\put(220.0,113.0){\rule[-0.200pt]{0.400pt}{140.686pt}}
\put(251,675){\raisebox{-.8pt}{\makebox(0,0){$\Diamond$}}}
\put(405,538){\raisebox{-.8pt}{\makebox(0,0){$\Diamond$}}}
\put(682,271){\raisebox{-.8pt}{\makebox(0,0){$\Diamond$}}}
\put(266,683){\raisebox{-.8pt}{\makebox(0,0){$\Diamond$}}}
\put(451,500){\raisebox{-.8pt}{\makebox(0,0){$\Diamond$}}}
\put(759,197){\raisebox{-.8pt}{\makebox(0,0){$\Diamond$}}}
\put(251,675){\usebox{\plotpoint}}
\put(241.0,675.0){\rule[-0.200pt]{4.818pt}{0.400pt}}
\put(241.0,675.0){\rule[-0.200pt]{4.818pt}{0.400pt}}
\put(405,538){\usebox{\plotpoint}}
\put(395.0,538.0){\rule[-0.200pt]{4.818pt}{0.400pt}}
\put(395.0,538.0){\rule[-0.200pt]{4.818pt}{0.400pt}}
\put(682.0,267.0){\rule[-0.200pt]{0.400pt}{2.168pt}}
\put(672.0,267.0){\rule[-0.200pt]{4.818pt}{0.400pt}}
\put(672.0,276.0){\rule[-0.200pt]{4.818pt}{0.400pt}}
\put(266.0,682.0){\usebox{\plotpoint}}
\put(256.0,682.0){\rule[-0.200pt]{4.818pt}{0.400pt}}
\put(256.0,683.0){\rule[-0.200pt]{4.818pt}{0.400pt}}
\put(451.0,493.0){\rule[-0.200pt]{0.400pt}{3.132pt}}
\put(441.0,493.0){\rule[-0.200pt]{4.818pt}{0.400pt}}
\put(441.0,506.0){\rule[-0.200pt]{4.818pt}{0.400pt}}
\put(759.0,183.0){\rule[-0.200pt]{0.400pt}{5.541pt}}
\put(749.0,183.0){\rule[-0.200pt]{4.818pt}{0.400pt}}
\put(749.0,206.0){\rule[-0.200pt]{4.818pt}{0.400pt}}
\put(266,683){\usebox{\plotpoint}}
\multiput(266,683)(14.781,-14.571){34}{\usebox{\plotpoint}}
\put(759,197){\usebox{\plotpoint}}
\end{picture}
\caption{Probabilities of $k=1,2$ and $3$ simultaneous clusters on 
lattices with periodic and free boundary conditions. $X_{PBC}=k^2-1/4$ 
and $X_{FBC}=k^2-k/2$, correspondingly.}
\label{fig:Pk} \end{figure}
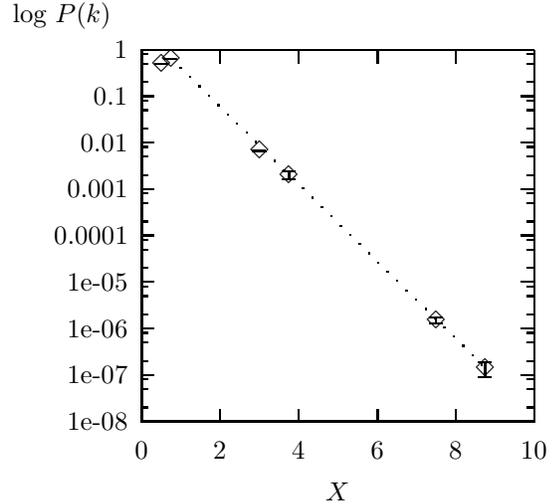
Therefore, using the known $k$-dependence of the probabilities, we can
collapse all but $k=1$ data on a single curve (Fig.\ref{fig:Pk}) versus
a Cardy variable $X$ which is $X_{PBC}=k^2-1/4$ for periodic boundaries and
$X_FBC=k^2-k/2$ for free boundaries.
The linear fit should give us an approximation of the exponent
$\alpha_{asymp}$ in (2), which is  $2\pi/3=2.094$ by Cardy's conjecture.
The linear fit to our data gives $1.91(20)$, which is, again, not so bad
an approximation to Cardy's asymptotic value.

It seems the fact that there are nonzero probability (about half 
of percent) of the coexistence of infinite clusters of black (open) and 
white (closed) sites (bonds) in critical two-dimensional percolation.

\end{document}